\documentstyle[psfig]{lamuphys}
\begin{document}

\title{FRACTAL GROWTH WITH QUENCHED DISORDER}
%\begin{document}
 \author{L. PIETRONERO \inst{1}, R. CAFIERO \inst{1} 
and A. GABRIELLI\inst{2}}
\institute{Dipartimento di Fisica, Universit\'a di 
Roma "La Sapienza", \\
P.le Aldo Moro 2, I-00185 Roma, Italy; Istituto Nazionale 
di Fisica della Materia, unit\'a di Roma I \and Dipartimento 
di Fisica, Universit\'a di Roma "Tor Vergata", \\
Via della Ricerca Scientifica 1, I-00133 Roma, Italy}

\maketitle

\begin{abstract}
In this lecture we present an overview of the physics of irreversible 
fractal growth process, with particular emphasis on a class of 
models characterized by {\em quenched disorder}.
 These models exhibit 
self-organization, with critical properties developing 
spontaneously, without the fine tuning of external parameters. This 
situation is different from the usual critical phenomena, and requires
 the introduction of new theoretical methods. Our approach to these 
problems is based on two concepts, the Fixed Scale Transformation,
 and the quenched-stochastic transformation, or 
Run Time Statistics (RTS), 
which maps a dynamics with quenched disorder into a stochastic 
process. These methods, combined together, allow us to 
understand the self-organized nature of models with quenched 
disorder and to compute analytically their critical exponents. In 
addition, it is also possible characterize mathematically the origin of 
the dynamics by {\em avalanches} and compare it with the {\em 
continuous growth} of other fractal models. A specific application to 
Invasion Percolation will be discussed. Some possible relations to 
glasses will also be mentioned.

\end{abstract}

\section{INTRODUCTION}
The introduction of the 
fractal geometry (\cite{mandel}) 
has changed the way physicists look at a vast class of natural 
phenomena which produce irregular structures. Many models have 
been introduced since the early eighties trying to relate these structures to well defined physical phenomena. These are 
the Diffusion Limited Aggregation 
(DLA) (\cite{dla}), the Dielectric Breakdown Model (DBM) (\cite{dbm}), 
the Invasion Percolation (IP) (\cite{invperc}), the Sandpile 
(\cite{btw}), the Bak and Sneppen 
model (BS) (\cite{bak}), just to give some examples. All these models 
lead spontaneously (for a broad range of parameters) 
to the development of critical properties and 
fractal structures. 

In the last years these has been a great interest on fractal 
models characterized by {\em quenched disorder}. These models 
are 
generally characterized by an intermittent dynamics, with bursts of 
activity of any size concentrated in a region of the system 
({\em avalanches}), and by memory effects induced by the presence of 
quenched disorder and by the dynamical rules 
(\cite{gap}; \cite{rtslung}; \cite{rrw}; \cite{matmem}). These 
memory effects in fractal growth processes 
with quenched disorder, may resemble an 
element characteristic of the spin glasses and glass dynamics. 
In fact, also in spin glasses, memory effects (aging) 
are due to the presence of quenched disorder, 
and are relied to the typical Kolrausch stretched 
exponential relaxation dynamics. However, at the moment it is hard
 to develop such an analogy in a more concrete way.

The study of physical phenomena leading to fractal structures 
can be classified by three different levels:
\begin{itemize}
\item{{\em Mathematical Level: Fractal Geometry}. This is a 
descriptive level, at which one simply recognizes the 
fractal nature of the phenomena and extimates the fractal dimension 
$D$.}
\item{{\em Physical Models}. One develops a model 
of fractal growth
 based on the physical process. This level is the analogue of 
the Ising model in equilibrium statistical mechanics.}
\item{{\em Physical Theories}. This level corresponds to a fully 
understanding of the origin of fractals in nature, their self-
organization etc. The corresponding level for phase transitions is 
the Renormalization Group.}
\end{itemize}
The analogy we make with phase transitions is quite natural, 
because, like ordinary critical phenomena, fractal growth 
models are scale invariant. However, some profound differences, 
like irreversible, non-equilibrium dynamics and SOC, make unavoidable the development of new theoretical concepts.

Fractal physical models can be classified into two main groups:
\begin{enumerate}
\item{Irreversible stochastic models}
\item{Irreversible quenched models}
\end{enumerate}
In the next section we discuss these models in relation with standard critical phenomena.
\section{PHYSICAL MODELS FOR SELF-SIMILAR GROWTH}

We briefly mention below some examples of these 
two classes of models, with a particular emphasis 
on models with quenched disorder.

1) Irreversible stochastic models

\begin{itemize}
\item{\em Diffusion Limited Aggregation (DLA)} (\cite{dla}). 
This is the first physical model of fractal growth. 
Particles performing a Brownian motion aggregate and 
form complex fractal structures.
\item{\em Dielectric Breakdown Model (DBM)} (\cite{dbm}). 
Is a generalization of DLA via the relation between potential 
theory and random walk.
\item{\em Sandpile Models} (\cite{btw}). These models are 
inspired by the marginal stability of sandpiles. The random 
addition of sand grains drives the system into a stationary 
state with a scale invariant distribution of avalanches.
\end{itemize}

2) Irreversible Quenched Models

\begin{itemize}
\item {\em Invasion Percolation (IP)} (\cite{invperc}). 
This model was 
developed to simulate the capillary displacement of a fluid in 
a porous medium. The porous medium is represented by a lattice 
where to each bond $i$ is assigned a quenched value $x_i$ 
of its conductance. At each time step the dynamics 
of the fluid evolves by occupying the 
bond with the smallest conductance between 
all its perimeter bonds. 
We call this kind of dynamics {\em extremal dynamics}.
IP is known to reproduce asymptotically the Percolation 
cluster of standard critical Percolation.

The main characteristics of this model are:
\begin{enumerate}
\item {\em Deterministic dynamics}. Once a 
realization of the quenched 
disorder is chosen, the dynamical rule selects 
in a deterministic way 
the bond to be invaded.
\item {\em Self-organization}. The process 
spontaneously develops 
scale-invariant structures and critical properties. In the limit $t \to 
\infty$ both long range space and time correlations appear.
\item {\em Avalanches}. The asymptotic dynamical 
evolution consists of 
local, scale-invariant macro-events, composed 
by elementary growth 
steps spatially and causally connected, called 
{\em avalanches}. When 
an avalanche stops, the activity is transferred 
to another region of the 
perimeter. 
\end{enumerate}
\item {\em Bak and Sneppen Model (BS)} (\cite{bak}). 
This model is similar 
to IP and has the same properties exposed in points 1-3. In fact, 
points 1-3 are characteristic of the whole 
class of SOC models with 
quenched disorder and extremal dynamics ( for a review 
see \cite{gap}). The BS model has 
been introduced to describe scale 
free events in biological evolution.

\item{\em Quenched models with an external modulating field}. This 
is a particular class of models, the prototype of which is the {\em 
Quenched Dielectric Breakdown Model (QDBM} (\cite{qdbm}; 
\cite{qdbm1}). This model 
is a sort of combination between IP (quenched) and DBM (stochastic). 
To each bond of a lattice is assigned a quenched 
random number $x_i$, representing the local resistivity. In addition, 
an external electric field $E$ is introduced. The ratio $y_i=x_i/E_i$ 
between quenched disorder and local electric field 
is the inverse of the 
current flowing in the bonds. At each time 
step the dynamics breaks 
the bond with the smallest $y_i$, that is to say with the biggest 
current. The QDBM modelizes the dielectric 
breakdown of a disordered 
solid. A similar model can be formulated to 
modelize the propagation 
of fractures in a inhomogeneous solid (\cite{qfrac}).
\end{itemize}

\begin{figure}
%\vspace{5.5cm}
%\epsfxsize 6.5cm
%\epsfysize 5.0cm
\psfig{figure=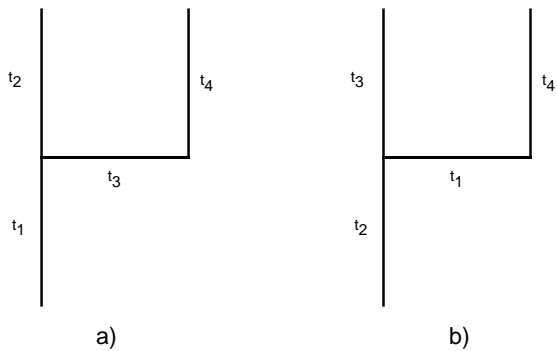,height=5.5cm}
\caption[]{For systems with irreversible dynamics, the statistical 
weight of a configuration depends on the whole growth history. 
The two identical configuration shown here have, in general, 
different statistical weights because their histories, indicated by
 the times $t_i\,\,\,i=1,2,3,4$, are different.}
\label{fig1}
\end{figure}

All the models in these classes share some 
important properties, that 
differentiates them from ordinary critical 
phenomena. First of all they 
are characterized by an {\em irreversible dynamics}, so 
that they cannot be described 
in hamiltonian terms, and the statistical weight of a configuration 
depends on the complete growth history (see. fig. \ref{fig1}). 
Quenched 
models have an additional problem, in that they have a 
deterministic 
dynamics, and the stochasticity enters only in the choice of the 
realization of the disorder. So, it is difficult to define transition 
probabilities for the dynamics.

Another fundamental difference, is that these models are self-
organized. Their dynamics evolves spontaneously 
in the phase space 
towards an {\em attractive} fixed point. No fine tuning of any 
parameter is needed. In addition, quenched models 
with extremal dynamics have an 
{\em avalanche dynamics}, that is to say the system 
tends to concentrate its 
activity in a well localized region of the perimeter, during an 
avalanche. When an avalanche stops, the 
activity transfers to another 
region of the perimeter and a new avalanche starts. 
On the contrary, in stochastic 
models, like DLA, there is a continuous growth process, 
that is to say for large systems the probability to have two nearby 
subsequent growth events tends to zero. The dynamical activity is 
diffused at each time on the whole growth interface. 

In table. \ref{table} we 
propose a scheme of comparison between the properties 
of ordinary critical phenomena and the 
most popular stochastic and quenched fractal growth models.

\section{NEW THEORETICAL CONCEPTS}

The application of the standard theoretical methods of 
statistical physics (field theory and renormalization group) 
is, in general, not possible for the main fractal growth problems, 
like DLA, DBM, Invasion Percolation, the sandpile model, 
which are characterized by an intrinsically irreversible dynamics. 

Here we discuss two theoretical methods we have developed 
in the past few years, as a step 
towards the construction of a physical theory for self-organized 
fractal growth processes. The exposition will be 
colloquial, and mainly devoted to the theoretical 
analysis of quenched 
models. For more details readers should refer to the bibliography.

\subsection{FIXED SCALE TRANSFORMATION (FST)}

This approach combines a technique of lattice path integral, to take 
into account the irreversible dynamics, with the study of the scale 
invariant dynamics inspired by the RG theory. It permits a 
description of the scale invariant properties of fractal growth 
\begin{table}
\caption[]{Comparison between the properties of ordinary critical 
phenomena represented by the Ising model and the most popular 
stochastic and quenched models generating fractal or scale 
invariant structures in a self-organized way.}
\psfig{figure=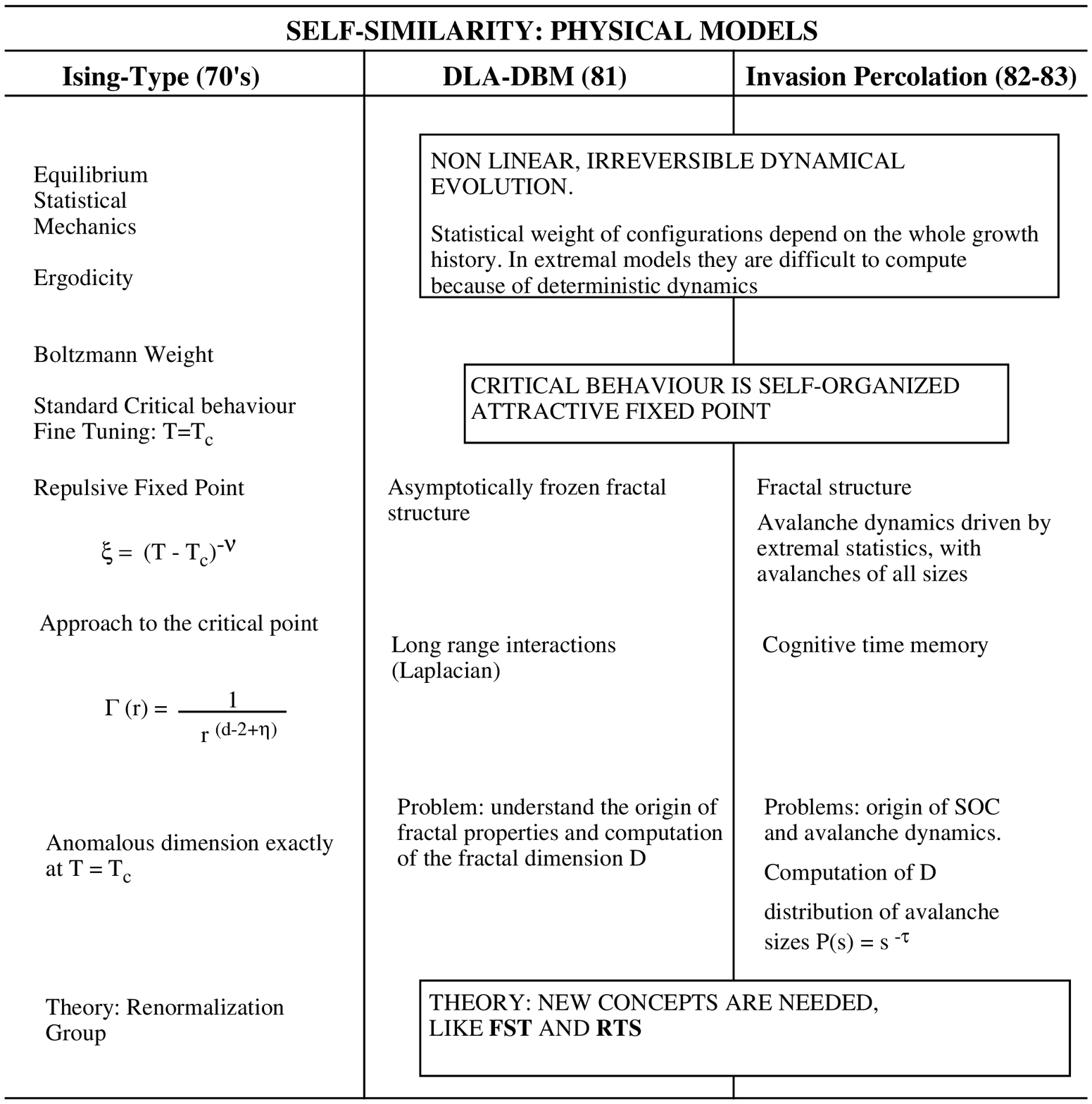,height=12.2cm}
%\vspace{9.0cm}
\label{table}
\end{table}
models. 

The method focuses on the dynamics at a given scale and 
analyzes the nearest neighbours correlations at this scale using 
{\em  lattice path integral} approach by which one can calculate the 
elements of a probability matrix, the FST matrix. The fixed point of 
the FST matrix gives the nearest neighbour correlations, at that 
scale. If one uses the scale invariant dynamics of the system, that 
can be obtained by Real Space Renormalization Group approaches 
(\cite{Raf}), one can generalize these correlations to all 
scales and compute the fractal dimension (\cite{FST2}). 

The 
basic point of the FST is the separation of the long time 
limit ($t \rightarrow \infty$) for the dynamical process at a given 
scale, from the large scale limit ($r \rightarrow \infty$), that 
defines the scale invariant dynamics. The interesting feature 
of FST is that it works at a fixed scale, so it is possible to include the 
fluctuation of the boundary conditions, that in systems with long 
range interactions, like {\em Diffusion Limited Aggregation} (DLA), 
have a great influence on the fractal properties. For this 
reason, the FST approach allows to reach a remarkable level of 
accuracy in the calculation of the 
fractal dimension. 

At the moment, the FST framework, eventually combined with the 
RTS method that we discuss below, seems to be the only general 
approach to understand the self-organized critical nature of a broad 
class of models going from DLA, to Percolation, to
 sandpile, to Invasion Percolation (\cite{FST2}; \cite{rtslung}). This 
situation therefore supports the idea that these models pose new 
questions for which one would like to develop a common theoretical 
scheme.

\subsection{QUENCHED-STOCHASTIC TRANSFORMATION (RTS)}

As we mentioned above, quenched models with extremal 
dynamics, like Invasion 
Percolation (IP), have a deterministic dynamics. This 
makes it impossible to address directly these class of model 
with the FST method or any other microscopic theory. Here
 we describe a general theoretical approach which addresses 
the basic problems of extremal models: 
(i) the understanding of the scale-invariance and 
self-organization;(ii) the origin of the avalanche
dynamics and (iii) the computation of 
the independent critical exponents. We will discuss in particular a 
specific application to Invasion Percolation (IP) (\cite{rtslung}), 
but these ideas can be easily extended to other models 
of this type like the Bak and Sneppen
model (\cite{BSRG}).

In order to overcome the problem represented by quenched 
disorder, we introduced a mapping of 
the quenched extremal dynamics into a stochastic 
one with cognitive memory, the {\em 
quenched-stochastic transformation}, also called
 Run Time Statistics (RTS) (\cite{RTS}). 
This approach was improved in various
steps (\cite{46matt}; \cite{46mattb}; \cite{RTS}; \cite{BSRG}) 
and now we can develop it into a general
theoretical scheme, that we call RTS-FST method (\cite{rtslung}). 
Its essential points are:
\\
- Quenched-stochastic transformation.
\\
- Identification of the microscopic fixed point
dynamics. This point clarifies the {\sl SOC nature} of the problem.
\\
- Identification of the scale invariant dynamics 
for block variables. This elucidates the origin of fractal structures.
\\
- Definition of {\em local} growth rules for the extremal model. This 
clarifies the origin of {\sl avalanche dynamics}.
\\
- Use of the above elements in a real space scheme, 
like the FST, to compute analytically the 
relevant exponents of the model.

 A general stochastic 
process is based on the following elements: a) a set
of time dependent dynamical 
variables $\{ \eta_{i,t} \}$; b) a Growth Probability Distribution 
(GPD) for the single growth step $\{ \mu_{i,t} \}$,
obtained from the $\{ \eta_{i,t} \}$; 
c) a rule for the evolution of the 
dynamical variables $\eta_{i,t}\to\eta_{i,t+1}$. 

Therefore, in order to map IP onto a stochastic process we have to: 
a) find the correct dynamical variables (the $\{ \eta_{i,t} \}$'s);
b) determine the GPD $\{\mu_{i,t}\}$ in terms of these variables;
c) find the evolution rule of the $\{ \eta_{i,t} \}$

A simple example can be useful to get an insight into the essence of 
the problem. Consider two independent random variables 
$X_1,X_2$, with uniform distribution $p_0(x_1)=p_0(x_2)=1$  in 
$[0,1]$ and 
let us eliminate the smallest, for example $X_2$. 
Clearly the probability
that $X_2<X_1$ is $1/2$. 
At the second ``time step'', we compare the
surviving variable $X_1$ with a third, uniform, random 
variable $X_3$ just added to the game and, again, we 
eliminate the smallest
one. At first sight one might think that, 
since both variables are independent, the probability
that $X_1$ survives again is $1/2$, but this is actually incorrect. 
In this case we indeed need to calculate the 
probability $\mu_3$ that $X_3<X_1$ \underline{given}
that $X_2<X_1$. This, using the rules of conditional 
probability, reads:
$$\mu_3=\tilde{P}(X_3<X_1)=P(X_3<X_1|X_2<X_1)=$$
\begin{equation}
 =\frac{P(X_3<X_1\bigcap X_2<X_1)}{P(X_2<X_1)}=\frac{2}{3},
 \label{eqexamp1}
\end{equation}
where $P(A|B)$ is the probability of
the event $A$, given that $B$ occurred, and 
$P(A\bigcap B)$ is the probability of occurrence of 
both $A$ and $B$. The point is that
the distribution of the variable $X_1$ is 
\underline{no longer} uniform when
it is compared with $X_3$, even though they are 
independent. The information that $X_2<X_1$ 
{\em changes in a conditional way} the {\em effective} 
probability density
$p_1(x)$ of $X_1$. Indeed the probability that
$x\le X_1<x+dx$ must now account for the fact that
$X_2<x$. By imposing this condition, we get: $p_1(x)=2x$. 
An analogous calculation for the 
distribution of $X_2$ gives $p_2(x)=2(1-x)$.
Qualitatively, the event $X_2<X_1$ decreases the
probability that $X_1$ has small values. On the
contrary, the probability that $X_2$ is small is
enhanced.

The above example contains the essential idea 
of the {\em quenched-stochastic transformation}. 
Extremal dynamics establishes, at each time step 
$t$, an order relation between quenched variables 
($X_2<X_1$ in the example). This {\em information} 
on the statistical properties of the variables 
involved in the process ({\em active} variables) 
can be conditionally stored 
in the form of their {\em effective densities}. 
Variables which have experienced the same 
dynamical history, will have the same effective 
density, irrespectively of their spatial position. 
This {\em memory} is represented 
by the {\em age} $k=t-t_0$, where $t$ is the actual 
time and $t_0$ is the time at which the variable 
became active. The effective 
densities $p_{k,t}(x)$ ($p_{0,t}(x)=p_{0,0}(x)=1$) 
of variables of age $k$ at 
time $t$ are {\em the 
dynamical variables of the stochastic process we are looking for}. 

A generalization of our simple example (Eq. \ref{eqexamp1}) 
leads to 
the following equation for the growth probability $\mu_{k,t}$ 
of a variable of age $k$ at time $t$ (GPD):
\begin{equation}
\mu_{k,t}
= \int_{0}^{1}dx\:p_{k,t}(x)\prod_{\theta} 
(1-P_{\theta,t}(x))^{n_{\theta,t}-\delta_{\theta,k}},
\label{mu}
\end{equation}
where $P_{\theta,t}(x)=\int_0^xdyp_{\theta,t}(y)$, the 
product is intended over all the ages of the active variables
and $n_{\theta,t}$ is the number
 of active variables of {\em age} $\theta$ at time $t$. The meaning
 of this expression is that the product inside the integral
takes into account of the competition of the
 selected variable with each of the other active variables.
The density $m_{k,t}(x)$ of this (smallest) 
variable after its growth 
is conditioned by the information that it has grown, and can
be computed from Eq. \ref{mu}. The temporal evolution 
of the densities of the still active 
variables is then given by:
\begin{equation}
p_{\theta+1,t+1}(x) = p_{\theta,t}(x) \int_{0}^{x}
\frac{m_{k,t}(y)}{1-P_{\theta,t}(y)}dy.
\label{nuova}
\end{equation}
Equations \ref{mu}, \ref{nuova} {\em
 accomplish our goal to describe a quenched extremal 
process as a stochastic process with time
memory}. The presence of memory is enlighted by the
 dependence of the GPD on the parameter $k$. A mean field like
expansion of Eq.(\ref{mu}) in the limit $t\to
\infty$ gives (\cite{RTS}): $\mu_{k,\infty}\sim 
\frac{1}{(k+1)^{\alpha}}$. This result is also confirmed by 
simulations (\cite{matmem}). Memory is at the origin of 
{\em screening effects} in the GPD $\{\mu_{k,t}\}$. The power
law behaviour of $\mu_{k,t}$ guarantees that
screening is preserved
 at all scales, which is the condition 
 to generate holes of all sizes in a growing pattern, 
leading to fractal structures (\cite{Raf}).

\begin{figure}
%\vspace{5.5cm}
%\epsfxsize 6.5cm
%\epsfysize 5.0cm
\psfig{figure=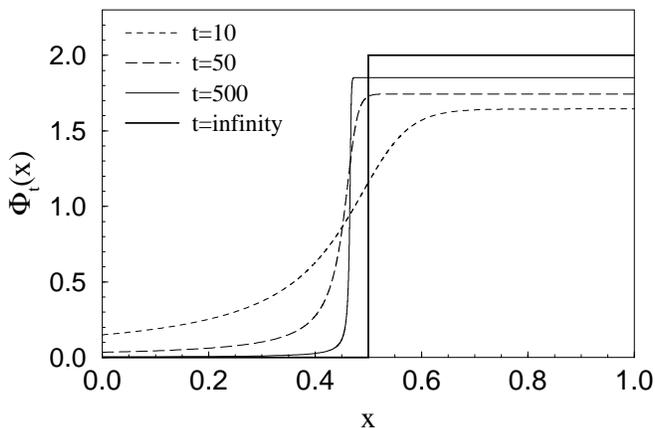,height=7.0cm}
\caption[]{Time evolution of the solution of the equation
for $\Phi_t(x)$ (for $2-d$ bond IP). $\Phi_t(x)$ 
tends asymptotically to a theta function with 
discontinuity at $p_c=1/2$.}
\label{figphi}
\end{figure}

This mapping, applied to models like IP, allows us to characterize 
mathematically the self-organization. In fact, the following {\em 
histogram equation} can be derived by the RTS equations \ref{mu}, 
\ref{nuova}:
 \begin{equation}
\partial_{x}\Phi_{t}(x)=\beta\Omega_{t}
\Phi_{t}^{2}(x)\:\left[1-\frac{\omega_t}
{\omega_t+1}\Phi_{t}(x)\right]
\label{eqphi}
\end{equation}
where $\omega_t=\langle{N_{t+1}-N_t}\rangle$, 
$\Omega_t=\langle{N_t}\rangle$, $N_t$ is the number 
of interface variables at time $t$, and $\beta$ is the 
solution of: $\beta=1-e^{-\beta \Omega_t}$. 
 This equation describes the time evolution of the distribution 
$\Phi_t(x)$ of quenched disorder on the growth interface.
The solution of eq. \ref{eqphi} becomes
 asymptotically (fig.\ref{figphi}) (\cite{RTS}): 
\begin{equation}
\lim_{t \rightarrow \infty}
\Phi_t(x)=\frac{1}{1-p_c}\theta(x-p_c)
\label{limphi}
\end{equation}
where $p_c$ is a critical threshold of the original extremal 
dynamics ($p_c=1/2$ for $2d$ bond IP), in agreement with 
numerical simulations (\cite{invperc}). Note that, in order to 
obtain the asymptotic behaviour \ref{limphi}, no fine tuning 
of any parameter is needed. This
clarifies the SOC nature of the problem.

The scale invariant dynamics can be shown to coincide with the 
microscopic one (\cite{rtslung}). The RTS approach permits also to 
characterize the origin of avalanche dynamics and to write down a 
set of equations describing the evolution of a single avalanche, by a 
straightforward modification of equations \ref{mu}, \ref{nuova} 
(\cite{rtslung}). From simulations (\cite{Maslov}), and from the 
histogram equation, one deduces that, asymptotically, each 
avalanche starts with a variable (initiator) equal to the 
threshold $p_c$ ($p_c=1/2$ for $2d$ bond IP). All other 
 variables in the avalanche 
 have values smaller than $p_c$. In view 
of the above arguments, the RTS equations
  for the {\em local avalanche dynamics} are obtained 
  from Eqs. (\ref{mu}, \ref{nuova})
  by taking into account only 
  the variables which become active after the initiator's 
  growth and by integrating in Eq.(\ref{mu})
  only in $[0,p_c]$. 

By using the equations for the local avalanche dynamics together 
with the FST method we have been able to compute with a very 
good accuracy (tipically $1-2 \%$, depending on approximations), 
the relevant critical exponents of 
Invasion Percolation, that is to say the fractal dimension and the
avalanche exponent (\cite{rtslung}). The method can be applied 
successfully also to the Bak and Sneppen model (\cite{BSRG}). 
In table \ref{table1} we show the theoretical values of 
the exponents of IP, that we 
have computed with the RTS-FST method, compared with numerical 
simulations

\begin{table}
\begin{centering}
\caption[]{Theoretical values of the fractal dimension of IP, with 
($D_f^{trap}$) and without trapping ($D_f$) and of directed IP 
($D_f^{DIP}$), and of the avalanche exponent $\tau$. These 
values are compared with numerical 
simulations.}
\label{table1}
\begin{tabular}{cccccc}
\hline
$ $ & $D_{f}$ & $D_{f}^{trap}$ 
& $D_{f}^{DIP}$ & $\tau^{IP}$ & $\tau^{trap}$ \\\hline\hline
$RTS-FST$ & $1.8879$ & $1.8544$ & $1.7444$ & $1.5832$ & 
$1.5463$ \\
\hline
$simul.$ & $\sim1.89$ & $\sim1.86$ & $\sim 1.75$ & $\sim 1.60$ & 
$\sim 1.53$\\
\hline
\end{tabular}
\end{centering}
\end{table} 
Recently, we have extended the RTS-FST scheme to QDBM, 
and we have obtained interesting, although preliminar, 
results, which allow us to elucidate some important characteristics of 
the class of models to which QDBM belongs (\cite{MQM}). 

\section{FURTHER DEVELOPMENTS}

In this lecture we have discussed two recently introduced theoretical 
methods, the FST and the RTS. These approaches have been applied sucesfully to many models for fractal growth, and allow to make a 
significative step towards the formulation of a common theoretical 
scheme for the physics of self-organized fractal growth. 

At the moment, we are studying the application of the RTS mapping to 
interface dynamics in quenched disorder (\cite{Snep}), and to glassy 
type dynamics. An interesting work, for what concerns the last point, 
is and RTS-type analysis of the statistical and dynamical properties of
 the random walk in quenched disorder (RRW) (\cite{rrw}), which has 
been studied by many authors as a toy model for localization 
(\cite{local}), depinning transitions (\cite{dep}), and aging effects 
(\cite{aging}). In this work, the authors map, by using 
the RTS method, the RRW dynamics into a 
stochastic dynamics with cognitive memory and recover all the 
characteristics of the original model. This suggests a link between 
stochastic dynamics with memory and the realizations of a dynamics 
with quenched disorder.

\end{document}